\documentstyle[11pt,epsf,paspconf]{article}

\begin{document}

\newcommand\hmpc{{h^{-1}\;{\rm Mpc}}}
\newcommand\hkpc{{h^{-1}\;{\rm kpc}}}

\title{Theoretical Modeling of the High Redshift Galaxy Population}

\author{David H. Weinberg}
\affil{Department of Astronomy, Ohio State University, Columbus, OH 43210}

\author{Romeel Dav\'e}
\affil{Astrophysical Sciences, Princeton University, Princeton, NJ 08544}

\author{Jeffrey P. Gardner}
\affil{Department of Astronomy, University of Washington, Seattle, WA 98195}

\author{Lars Hernquist}
\affil{Department of Astronomy, Harvard University, Cambridge, MA 02138}

\author{Neal Katz}
\affil{Department of Physics and Astronomy, University of Massachusetts,
Amherst, MA 98195}

\begin{abstract}
We review theoretical approaches to the study of galaxy formation, with
emphasis on the role of hydrodynamic cosmological simulations in
modeling the high redshift galaxy population.  We present new predictions
for the abundance of star-forming galaxies in the LCDM model (inflation $+$
cold dark matter, with $\Omega_m=0.4$, $\Omega_\Lambda=0.6$), combining
results from several simulations to probe a wide range of redshift.
At a threshold density of one object per arcmin$^2$ per unit redshift, these
simulations predict galaxies with star formation rates of $2M_\odot/$yr
($z=10$), $5M_\odot$/yr ($z=8$), $20M_\odot$/yr ($z=6$), $70-100M_\odot$/yr
($z=4-2$), and $30M_\odot$/yr ($z=0.5$).
For galaxies selected at a fixed comoving space density 
$n=0.003\;h^3{\rm Mpc}^{-3}$, a simulation of a $50\hmpc$ cube predicts
a galaxy correlation function $(r/5\hmpc)^{-1.8}$ in comoving coordinates,
essentially independent of redshift from $z=4$ to $z=0.5$.  Different 
cosmological models predict global histories of star formation that reflect
their overall histories of mass clustering, but robust numerical predictions
of the comoving space density of star formation are difficult because the
simulations miss the contribution from galaxies below their resolution limit.
The LCDM model appears to predict a star formation history with roughly the
shape inferred from observations, but it produces too many stars at low
redshift, predicting $\Omega_\star\approx 0.015$ at $z=0$.  We conclude with
a brief discussion of this discrepancy and three others that suggest gaps in
our current theory of galaxy formation: small disks, steep central halo
profiles, and an excess of low mass dark halos.  While these problems could 
fade as the simulations or observations improve, they could also guide us
towards a new understanding of galactic scale star formation, the spectrum of
primordial fluctuations, or the nature of dark matter.
\end{abstract}


\section{Theoretical Approaches to Galaxy Formation}

In broad outline, the current theory of galaxy formation is remarkably
similar to the one described by White \& Rees (1978) two decades ago.
Gravitational instability of primordial density fluctuations leads to
the collapse of dark matter halos.  Gas falls into these potential wells,
heats up as it does so, then radiates its energy, loses pressure support,
contracts, and eventually forms stars.
Dissipation thus leads to the formation of dense baryonic cores, which
can survive as distinct entities even if their parent dark halos merge.

Relative to the situation in 1978, we now have much better theoretical
models for the origin of the primordial fluctuations, with inflation
as the leading candidate.  There has also been a major change to the
theoretical picture, the idea that the dominant mass component
is not baryonic dark matter but some form of non-baryonic, cold dark matter
(CDM).  No existing models that assume purely baryonic matter can
account for the cosmic structure seen today while remaining consistent
with the observed low amplitude of cosmic microwave background anisotropies.
In inflation+CDM models normalized to the COBE observations,
the properties of the initial conditions for structure
formation are completely determined by a small number of parameters,
principally $\Omega_m$, $\Omega_b$, and $\Omega_\Lambda$ (the density
parameters of matter, baryons, and vacuum energy), $H_0$, and the
inflationary spectral index $n$ (where $n=1$ corresponds to scale-invariant
fluctuations).  

There have also been substantial developments in the ``technology''
for modeling galaxy formation theoretically.  One line of attack,
semi-analytic modeling, descends directly from the approach of 
White \& Rees (1978).  These methods use extensions of the 
Press-Schechter (1974) formalism that describe merger histories
of collisionless dark matter halos (Bond et al. 1991, Bower 1991,
Lacey \& Cole 1994).  They adopt an idealized description of gas dynamics 
and cooling within halos and incorporate parametrized models of star
formation from cooling gas and reheating by supernova feedback
(e.g., White \& Frenk 1991, Kauffmann et al. 1993, Cole et al. 1994, 
Avila-Reese et al.\ 1998, Somerville \& Primack 1999).
A simpler branch of this work focuses on the properties of galaxy disks,
with the halo spin parameter playing a critical role
(Fall \& Efstathiou 1980, Dalcanton et al. 1997, Mo et al. 1998).
The semi-analytic approach allows traditional population synthesis
and chemical evolution models to be placed in a far more realistic
framework of structure formation.  There are numerous free parameters
to be set either by normalization to observations or by calibration
against numerical simulations, but once these parameters are fixed
the models can be tested against many independent observations.
The strengths of semi-analytic models lie in their ability to make
contact with a wide range of observations, their ability to explore
a wide range of theoretical parameter space,
and their description of galaxy formation in
simple and physically intuitive terms.  The semi-analytic approach
can also be combined with N-body simulations to sidestep some of the
approximations used in the halo merger models and, more importantly,
to obtain detailed predictions of galaxy clustering
(e.g., Governato et al. 1998, Kauffmann et al. 1999ab).

The other main approach to theoretical modeling of galaxy formation
is direct numerical simulation, including hydrodynamics and star formation.
There are two broad classes of simulations, those that zoom in on 
individual objects (e.g., Katz \& Gunn 1991,
Katz 1992, Navarro \& White 1994,
Vedel et al.\ 1994, Steinmetz \& M\"uller 1994,
Navarro \& Steinmetz 1997, Dominguez-Tenreiro et al.\ 1998, Kaellender \&
Hultman 1998) and those that simulate larger, random realizations of 
cosmological volumes (e.g., Cen \& Ostriker 1992, Katz et al.\ 1992, 
Evrard et al.\ 1994, Pearce et al.\ 1999).  The main goal of the first
class of simulations is to study the formation mechanisms and 
properties of individual galaxies, while the main goal of the second is
to study spatial clustering of the galaxy population and the statistical
distributions of galaxies' stellar, gas, and total masses.
There are two main technical approaches, Eulerian grid codes
and smoothed particle hydrodynamics (SPH), with Lagrangian grid codes
(Gnedin 1995, Pen 1998) having intermediate properties.
Eulerian grid codes can often achieve high mass resolution (i.e.,
have many grid cells and hence small mass per cell), but the uniformity
of the fixed grid means that the spatial resolution in computing 
hydrodynamic forces is usually rather low (though it can be improved
by using multiple grids, as in Anninos \& Norman [1996]).
For example, the
recent simulations of Cen \& Ostriker (1999) use 200$h^{-1}$ kpc
grid cells and attempt to identify sites and rates of galactic
level star formation based on the gas properties averaged over this scale.
SPH simulations compute hydrodynamic forces by smoothing over a fixed
number of particles and therefore have higher spatial resolution in
denser regions.  In SPH simulations that include radiative cooling,
the gas in well resolved halos almost invariably has a two-phase
structure, with dense, cold lumps embedded in a hot, pressure-supported
medium.  The cold lumps have sizes and masses roughly comparable to the 
luminous regions of observed galaxies, and they stand out distinctly
from the background, so that there is no ambiguity in identifying
the ``galaxies'' in such simulations.

Relative to semi-analytic modeling of galaxy formation,
the strength of the numerical simulation approach is its
more realistic treatment of gravitational collapse, mergers,
and heating and cooling of gas within dark halos.  
The difference in treatment could be quantitatively important,
since the simulations show that gas and galaxies are distributed along
filamentary networks that resemble the root system of a tree,
making the accretion process very different
from the spherically symmetric one envisaged in the semi-analytic
calculations.  The only free 
parameters (apart from the physical parameters of the cosmological
model being studied) are those related to the treatment of star 
formation and feedback.  Given these parameters, simulations provide
straightforward, untunable predictions.  However, the simulation
approach must contend with the numerical uncertainties caused by
finite volume and finite resolution, and computational expense
makes it a slow way to explore parameter space.

\section{Numerical Simulations of the High-Redshift Galaxy Population}

Our group's approach to hydrodynamic cosmological simulations is described
in detail by Katz et al.\ (1996).  We use a cosmological
version of the Hernquist \& Katz (1989) TreeSPH code;
the results shown here are from simulations that use the parallel
code developed by Dav\'e et al.\ (1997).
In brief, the simulations use two sets of particles to represent the
dark matter and gas components, respectively.  Dark matter particles
respond only to gravitational forces, which are computed using a
hierarchical tree method (Barnes \& Hut 1986).  Gas particles respond
to gravity and to gas dynamical forces computed by SPH.
The simulations include heating by shocks, adiabatic compression, and
photoionization, and cooling by adiabatic expansion,
Compton interaction with the microwave background, and, most importantly,
all of the radiative atomic processes that arise in a primordial
composition gas.

In simulations without star formation, a fraction of the gas condenses
into cold, dense lumps, with typical sizes of one to several kpc, and
masses up to a few $\times 10^{11} M_\odot$.
Our star formation algorithm is essentially a prescription for turning
this cold, dense gas into collisionless stars, returning energy
from supernova feedback to the surrounding medium.
A gas particle is ``eligible'' to form stars if it
is Jeans unstable, resides in a region of converging flow,
and has a physical density exceeding 0.1 hydrogen atoms cm$^{-3}$.
Once a gas particle is {\it eligible} to form stars,
its star formation {\it rate} is given by
\begin{equation}
{d\rho_\star \over dt} = - {d\rho_g \over dt} =
       {c_\star\epsilon_\star\rho_g \over t_g},
\label{eqn:sfrate}
\end{equation}
where $c_\star=0.1$ is a dimensionless star formation rate parameter,
$\epsilon_*=1/3$ is the fraction of the particle's gas mass that will
be converted to stellar mass in a single simulation timestep,
and the gas flow timescale $t_g$ is the maximum of the local gas dynamical
time and the local cooling time.
Recycled gas and supernova feedback energy are distributed to the
particle and its neighbors.  Because the surrounding medium is dense
and has a short cooling time, the feedback energy is usually radiated
away rather quickly, so it has only a modest impact in our simulations.

This description of galactic scale star formation is clearly 
idealized, but our predictions of galaxy properties are not sensitive
to its details.  As shown in Katz et al.\ (1996), changing $c_\star$
by an order of magnitude makes almost no difference
to the stellar masses of simulated galaxies.  With lower $c_\star$,
gas simply settles to higher density before star formation begins
to deplete it, and the dependence of $d\rho_\star/dt$ on $\rho_g$
and $t_g$ in equation~(\ref{eqn:sfrate}) ensures that star formation
cannot get too far out of step with the rate at which gas cools
out of the hot halo.  One could, however, imagine radically different
formulations of galactic scale star formation that would lead to 
different results, e.g., if cooled gas does not
form stars steadily but instead accumulates until an interaction
triggers a violent starburst.
One could also imagine a picture in which multi-phase
substructure in the ISM allows supernova feedback to have a greater
impact on the surrounding halo gas.

We have analyzed the clustering of galaxies at $z=2-4$ in simulations
of a variety of CDM models in Katz et al.\ (1999, hereafter KHW), 
and we will discuss
the star formation properties of the galaxies in those simulations
in a forthcoming paper.  Here we focus instead on results that cover 
a wider range of redshifts for a single model: flat cosmology with
$\Omega_m=0.4$, $\Omega_\Lambda=0.6$, $\Omega_b=0.0473$, $h=0.65$, $n=0.95$,
with a COBE-normalized mass fluctuation amplitude $\sigma_8=0.80$ on the
scale of $8\hmpc$ at $z=0$.  We have been carrying out simulations of
this LCDM model with a range of box sizes, particle numbers, and ending
redshifts, to address a variety of questions, including the influence
of numerical parameters on the predictions of high redshift structure.

\begin{figure}
\centerline{
\epsfysize=6.0truein
\epsfbox[25 58 565 740]{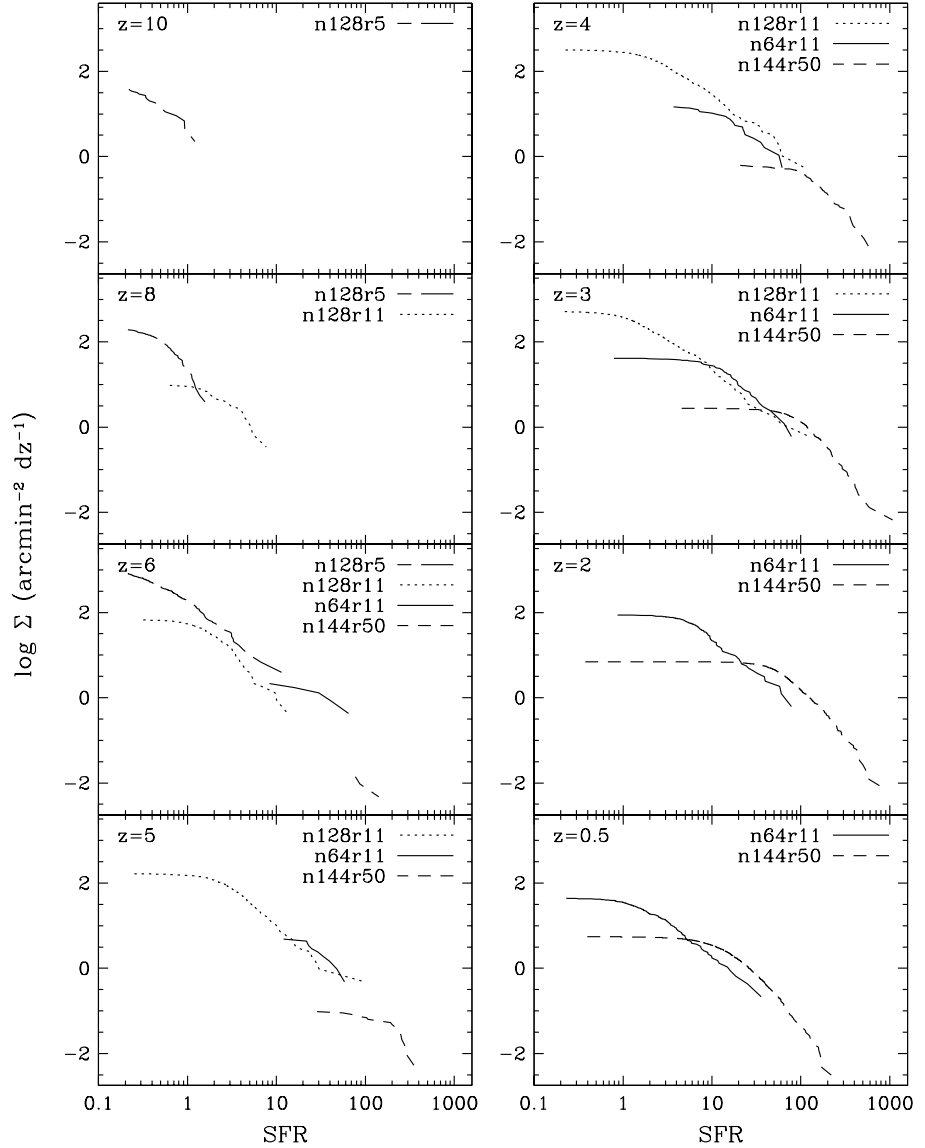}
}
\caption{The cumulative distribution of galaxies as a function of
instantaneous star formation rate, from $z=10$ to $z=0.5$,
in simulations of the LCDM model with different numerical parameters.
Each curve shows the surface density of simulated galaxies, in number per
arcmin$^2$ per unit redshift, with star formation rate above the
value indicated on the $x$-axis.  Legends indicate the simulation
box size (5.55, 11.11, or 50$\hmpc$) and particle number
($64^3$, $128^3$, or $144^3$ particles of each species).  
} \label{fig-1}
\end{figure}

Figure~\ref{fig-1} presents the result most relevant to those searching
for high-$z$ galaxies: the predicted surface density of objects as a
function of star formation rate (SFR), from $z=10$ to $z=0.5$.
The SFR can be converted approximately to a UV continuum luminosity
using one's favorite IMF and population synthesis model.  
For example, the model adopted by Steidel et al.\ (1996) gives
an absolute magnitude $M_{AB}=-20.8$ at rest-frame wavelength 1500\AA\
for SFR=$10 M_\odot/$yr.  
Each line type in Figure~\ref{fig-1} corresponds to a simulation
with a different number of particles and/or simulation box size, as
indicated by the legends.  In general, simulations with higher mass
resolution better represent the low end of the
luminosity function, while simulations with larger volume 
better represent the high end.  Since limited resolution and limited
box size both cause underestimates of galaxy numbers, one can generally
take the highest line at a given SFR in a given panel as a lower limit
to the predicted surface density of objects.  In panels where
the curves from different simulations connect up, we can infer that the
prediction based on the upper envelope of these curves is reasonably
robust to the numerical parameters of the simulations, though of course
it may still be sensitive to the cosmological parameters and to the
adopted model of star formation.

\begin{table}
\caption{Parameters of the LCDM Simulations} \label{tbl-1}
\begin{center}\scriptsize
\begin{tabular}{rrrrrr}
Title & $N$ & $R$ & $\epsilon_{\rm grav}$ & $60m_{\rm dark}$ & $60 m_{\rm SPH}$
\\
\tableline
n128r5	 &$128^3$   &5.55   &1.25 &$7.4\times 10^8$     &$9.9\times 10^7$ \\
n128r11	 &$128^3$   &11.11  &2.5  &$5.9\times 10^9$     &$7.9\times 10^8$ \\
n64r11	 &$64^3$    &11.11  &5    &$4.7\times 10^{10}$  &$6.3\times 10^9$ \\
n144r50	 &$144^3$   &50.0   &10   &$3.8\times 10^{11}$  &$5.1\times 10^{10}$ \\
\end{tabular}
\end{center}
\end{table}

Table~\ref{tbl-1} lists the parameters of the simulations shown in 
Figure~\ref{fig-1}:  the number of particles of each species,
the size of the simulation cube (in comoving $\hmpc$), the gravitational
force softening length (in comoving $\hkpc$), and approximate mass
resolution limits (in $M_\odot$).
In examination of the simulations, we find that dark 
halos containing at least 60 dark matter particles essentially always
contain a simulated galaxy, and we therefore list $60 m_{\rm dark}$
as a minimum resolved dark halo mass.  The requirement for robustly
estimating the SFR via equation~(\ref{eqn:sfrate}) seems to be somewhat
more stringent, 60 or more particles in the dissipated baryon component
({\it cold} gas $+$ stars), and we therefore list $60m_{\rm SPH}$ as
the minimum baryon mass of a resolved galaxy.  The curves in Figure~\ref{fig-1}
are computed only for galaxies above this resolution limit, and the
flat tails of some of these curves at low SFR suggest that even the
$60m_{\rm SPH}$ criterion may be a bit too loose.
Although we compute each galaxy's ``instantaneous'' SFR using 
equation~(\ref{eqn:sfrate}), we find that the SFR is reasonably
steady over timescales of 200 Myr, and our 
results would therefore not be very different if we averaged over
time intervals up to this value.

\begin{figure}
\centerline{
\epsfysize=2.5truein
\epsfbox[60 475 550 725]{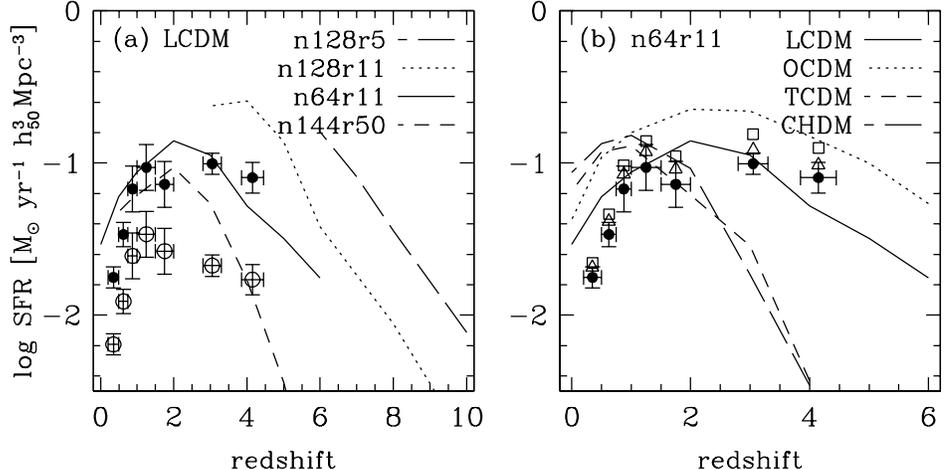}
}
\caption{Globally averaged star formation rates as a function of
redshift in {\it (a)} simulations of the LCDM model with various
numerical parameters, and {\it (b)} simulations of various CDM models,
all with $64^3$ particles in an $11.11\hmpc$ box.  Note the different
ranges of the $x$-axis in the two panels.  Data points are taken from
the compilation by Steidel et al.\ (1999).  In {\it (a)}, filled and
open circles show estimates with and without extinction corrections,
corrected to the LCDM cosmological parameters.  In {\it (b)}, 
extinction-corrected estimates are shown for the cosmological parameters
of LCDM (filled circles), OCDM (open triangles), and 
TCDM/CHDM (open squares).
} \label{fig-2}
\end{figure}

Figure~\ref{fig-2}a shows the average comoving density of star formation as
a function of redshift, a representation made famous by Madau et al.\ (1996),
for the same simulations plotted in Figure~\ref{fig-1}.
Of course, a given simulation only includes the contribution 
of galaxies above the resolution limit listed in Table~\ref{tbl-1}. 
At any given redshift, the higher resolution simulation always has
a higher comoving density of star formation, and the difference
between simulations generally grows with increasing redshift because
the contribution of low mass galaxies is more important at higher $z$.
Thus, the highest line at a given redshift in Figure~\ref{fig-2}a
should be taken as a lower limit to the predicted SFR.
It is tempting to draw an envelope that connects the tops of
the long-dashed, dotted, and solid curves, but it is not clear that
even this combined result is numerically converged, especially at $z>5$.

The data points in Figure~\ref{fig-2}a, kindly provided by C. Steidel,
are taken from the compilation
by Steidel et al.\ (1999), based on their own data ($z>3$) and the
data of Lilly et al.\ (1996, $z<1$) and Connolly et al.\ (1997; $1<z<2$).
Filled circles include Steidel et al.'s estimated corrections for
extinction of the rest-frame UV continuum, a factor of 2.75 at $z<2$
and 4.68 at $z>2$.  Open circles have no extinction correction.
We have converted all values to our $\Omega_m=0.4$, $\Omega_\Lambda=0.6$
cosmology.

Given that higher numerical resolution should only increase the predicted SFR,
it appears that the LCDM simulations predict more star formation than
is inferred from UV continuum observations, at least with Steidel et al.'s 
(1999) extinction estimates.  Given the uncertainties in the 
observational estimates (which depend on assumed IMF shapes and population
synthesis models in addition to extinction corrections), it is
premature to draw strong conclusions from this comparison.
However, the impression of excessive star formation is also supported by
an independent comparison at $z=0$.  The n64r11 simulation predicts
a density parameter $\Omega_\star = 0.0146$ in stars at $z=0$.
This is roughly four times higher than Fukugita et al.'s (1998)
observational estimate, $\Omega_\star = 0.0038^{+0.0024}_{-0.0017}$.
If the simulations are to reproduce observational estimates of $\Omega_\star$
at $z=0$, then a large fraction of the gas that is cooling into galaxies
must actually be forming baryonic dark matter rather than luminous stars.
Katz et al.\ (1992) reached a similar conclusion for an $\Omega_m=1$
CDM model, and Pearce et al.\ (1999) reached a similar conclusion 
for LCDM with an independent simulation.  

Figure~\ref{fig-2}b shows global star formation rates for the 
four CDM simulations analyzed in Dav\'e et al.'s (1998) study
of the low-redshift Ly$\alpha$ forest: a flat model with a cosmological
constant (LCDM), an open universe model with $\Omega_m=0.5$ (OCDM),
and two $\Omega_m=1$ models, one with a tilted inflationary spectrum
($n=0.8$, TCDM), and one with a hot dark matter component
($\Omega_\nu=0.2$, CHDM).  All of these simulations use $64^3$
particles of each species in an $11.11\hmpc$ box; the LCDM simulation
is the same as the one labeled n64r11 in Figures~\ref{fig-1} and~\ref{fig-2}a.
The differences in predicted star formation histories clearly 
reflect the differences in the overall history of mass clustering
in the various models.  The open model has the earliest structure
formation, and its star formation history peaks at the highest
redshift and shows the most rapid decline between $z=1$ and $z=0$.
The $\Omega_m=1$ models, on the other hand, form structure relatively
late, and their global star formation rates decline only mildly
at low redshift.  Because of their low mass fluctuation amplitudes
and relatively late structure formation, the TCDM and CHDM models
appear to be in serious conflict with the observationally estimated
star formation rates at $z>3$.  However, this conclusion could
be sensitive to the finite numerical resolution of the simulations,
since the main contribution to the global SFR in these models 
will come from galaxies below our resolution limit.

\begin{figure}
\centerline{
\epsfysize=2.5truein
\epsfbox[105 425 460 720]{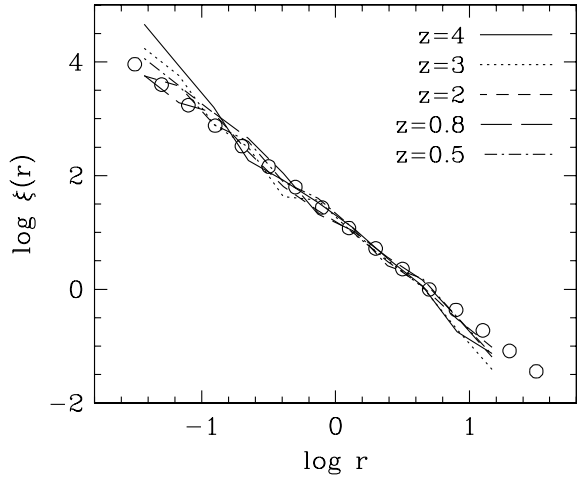}
}
\caption{Galaxy correlation functions in the LCDM model as 
a function of redshift, with $r$ in comoving $\hmpc$,
from the n144r50 simulation.
At each redshift, the correlation function is measured for the
365 most massive galaxies in the box.
Open circles show a power-law, $\xi(r) = (r/5\hmpc)^{-1.8}$.
} \label{fig-3}
\end{figure}

One of the striking features of the observed Lyman-break galaxy (LBG)
population is its high clustering amplitude, similar to that of
bright galaxies at $z=0$ (Adelberger et al.\ 1998; Giavalisco et al.\ 1998;
Steidel et al.\ 1998).  This strong clustering has a straightforward
theoretical explanation, analogous to the one given by Kaiser (1984)
for the strong clustering of Abell clusters at $z=0$.
Like clusters today, LBGs are ``rare'' objects relative to the scale
of non-linear mass clustering (at $z=3$), and as a result they are
highly biased tracers of the underlying mass distribution.
Numerical or analytic models that associate observed LBGs with 
massive dark halos can therefore explain their observed clustering
amplitude without much difficulty.

KHW found that the clustering of high redshift galaxies was remarkably
insensitive to the adopted cosmological model, because differences in
bias mask differences in clustering of the underlying mass distributions.
Analytic models lead to a similar conclusion (e.g., Adelberger et al.\ 1998).
The small volume of the KHW simulations ($11.11\hmpc$ cubes) made a
direct comparison to existing spectroscopic samples of LBGs difficult,
so in Figure~\ref{fig-3} we show results for the LCDM model from our
n144r50 simulation.  At each redshift we select the 365 most massive
galaxies (which is the full population at $z=4$ and a decreasing fraction at
lower $z$), so that we measure the clustering of a sample of fixed
comoving space density, $n=0.0029\;h^3{\rm Mpc}^{-3}$ (0.2 galaxies per
arcmin$^2$ per unit redshift at $z=3$).  Although the dark matter clustering
grows steadily from $z=4$ to $z=0.5$, the galaxy correlation function
stays virtually constant, well described at all redshifts by a power law
$\xi(r)=(r/5\hmpc)^{-1.8}$.  This result agrees well with that of
Pearce et al.\ (1999), who use a similar numerical technique. 
Cen \& Ostriker (1999), using a very different numerical technique,
find nearly identical values for the comoving correlation length
but a steeper correlation function slope.

\section{Some Outstanding Issues}

Hydrodynamic simulations of models like LCDM seem to provide a fairly
good account of galaxy formation.  Standard gas dynamics and radiative
cooling combined with plausible recipes for star formation lead to
objects with roughly the right sizes and masses to represent the
luminous regions of observed galaxies.  Simulations easily reproduce
the observed abundance of Lyman-break galaxies, and the global star
formation history that they predict has at least roughly the shape
suggested by observations.  The simulated galaxy populations also
exhibit approximately the correct clustering strength at
$z=0$ and at $z=3$.  Semi-analytic models can claim similar
successes (e.g., Baugh et al.\ 1998, Governato et al.\ 1998,
Benson et al.\ 1999, Kauffmann et al.\ 1999ab, Somerville et al.\ 1999),
though it is not yet clear whether the agreement of end results
between the two approaches arises (as one would hope) because both
are correctly modeling the same underlying physics.
Indeed, two semi-analytic models that adopt quite different scenarios
for the nature of Lyman-break galaxies can both reproduce their
observed properties fairly well (Baugh et al.\ 1998, Somerville et al.\ 1999).

An optimist would certainly conclude that the glass containing our
current theoretical understanding of galaxy formation is at least half full.
However, there are some nagging discrepancies that should not escape
mention, since they may well turn out to have interesting implications.
First is the problem already mentioned in \S 2, a tendency for hydrodynamic
simulations to process too many baryons into stars.  Baryonic dark matter
is an acceptable solution to this problem, but it is discomforting to 
be forced into such a dodge.  A second difficulty arises in simulations
of individual galaxies: lumpy, dissipative collapse causes baryons to transfer
angular momentum to the dark matter halo, with the result that the
final galaxy disks are too small (Navarro \& White 1994, Navarro \& Steinmetz
1997).  The third and fourth problems appear in high-resolution,
collisionless N-body simulations (no gas dynamics), which predict
dark matter halos that are too concentrated to match observed rotation curves
(Moore 1994, Flores \& Primack 1994, Moore et al.\ 1999b, 
Navarro \& Steinmetz 1999,
but see Kravtsov et al.\ 1998),
and an abundance of low mass satellite galaxies far in excess of the
number observed in the Local Group (Klypin et al.\ 1999, Moore et al.\ 1999a).

Each of these apparent discrepancies rests on a handful of technically
challenging numerical simulations, and one or more of them might
fade as the simulations (or the observational data)
improve.  However, it seems likely that at least some of these are
genuine physical problems.  The solutions may lie in the astrophysics
of galactic scale star formation --- in particular, strong supernova
feedback is often invoked as a way to reduce the overall level of 
star formation, delay gas cooling so as to suppress angular momentum loss,
prevent star formation in low mass satellite halos, and perhaps even rearrange
the mass in galaxy cores enough to change rotation curve shapes 
(Navarro et al.\ 1996).  Alternatively, the solutions may require
changes to the fundamental tenets of the cosmological models,
such as the spectrum of primordial fluctuations or the assumption
that dark matter is cold and non-interacting.  There is enough commonality
to the problems to suggest that they may well have a common set of solutions.
Comparison between increasingly accurate theoretical calculations
and the growing web of observational constraints at all redshifts
should eventually tell us whether those solutions lie in gas dynamics,
star formation, early universe physics, or the nature of dark matter.

\end{document}